\documentclass[pre,aps,10pt,a4paper,nofootinbib, superscriptaddress,onecolumn,english,floatfix]{revtex4-1}
\usepackage{babel}
\usepackage[utf8]{inputenc}
\usepackage[utf8]{luainputenc}
\usepackage[T1]{fontenc}
\usepackage{amsthm}
\usepackage{amsmath}
\usepackage{amssymb}
\usepackage{bm}
\usepackage{graphicx}
\usepackage{natbib}
\usepackage[]{units}

\usepackage[usenames]{color}
\usepackage{ulem}

\begin{document}
\title{Numerical simulations of self-diffusiophoretic colloids  at fluid interfaces}
\author{T. Peter}
\affiliation
{
   Max Planck Institute for Intelligent Systems, 
   Heisenbergstr.\ 3,
   70569 Stuttgart,
   Germany, 
}
\affiliation{
   Institute for Theoretical Physics IV,
   University of Stuttgart,
   Pfaffenwaldring 57,
   70569 Stuttgart,
   Germany
}
\author{P. Malgaretti}
\email{malgaretti@is.mpg.de}
\affiliation
{
   Max Planck Institute for Intelligent Systems, 
   Heisenbergstr.\ 3,
   70569 Stuttgart,
   Germany, 
}
\affiliation{
   Institute for Theoretical Physics IV,
   University of Stuttgart,
   Pfaffenwaldring 57,
   70569 Stuttgart,
   Germany
}
\author{N. Rivas}
\affiliation{Helmholtz Institute Erlangen-N\"urnberg for Renewable Energy (IEK-11), Forschungszentrum J\"ulich, F\"urther Stra$\beta$e 248,
90429 N\"urnberg, Germany}
\author{A. Scagliarini}
\affiliation{Helmholtz Institute Erlangen-N\"urnberg for Renewable Energy (IEK-11), Forschungszentrum J\"ulich, F\"urther Stra$\beta$e 248,
90429 N\"urnberg, Germany}
\affiliation{CNR-IAC, Institute for Applied Mathematics 'M. Picone', Via dei Taurini 19, 00185 Rome, Italy}
\author{J. Harting}
\affiliation{Helmholtz Institute Erlangen-N\"urnberg for Renewable Energy (IEK-11), Forschungszentrum J\"ulich, F\"urther Stra$\beta$e 248,
90429 N\"urnberg, Germany}
\affiliation{Department of Applied Physics, Eindhoven University of Technology, P.O. box 513, NL-5600MB Eindhoven,
The Netherlands
}
\author{S. Dietrich}
\affiliation
{
   Max Planck Institute for Intelligent Systems, 
   Heisenbergstr.\ 3,
   70569 Stuttgart,
   Germany, 
}
\affiliation{
   Institute for Theoretical Physics IV,
   University of Stuttgart,
   Pfaffenwaldring 57,
   70569 Stuttgart,
   Germany
}
\begin{abstract}
    The dynamics of active colloids is very sensitive to the presence of boundaries and interfaces which therefore can be used to control their motion. 
    Here we analyze the dynamics of active colloids adsorbed at a fluid-fluid interface. By using a mesoscopic numerical approach which relies on  an approximated numerical solution of the Navier-Stokes equation, we show that when adsorbed at a fluid interface, an active colloid experiences a net torque even in the absence of a viscosity contrast between the two adjacent fluids. In particular, we study the dependence of this torque on the contact angle of the colloid with the fluid-fluid interface and on its surface properties. We rationalize our results via an approximate approach which accounts for the appearance of a local friction coefficient.
    By providing insight into the dynamics of active colloids adsorbed at fluid interfaces, our results are relevant for two-dimensional self assembly and emulsion stabilization by means of active colloids.
\end{abstract}
\date{\today}
\maketitle
\section{Introduction}

The dynamics of synthetic or biological, self-propelled objects is strongly affected by the presence of boundaries and interfaces~\cite{Ebbens2010,Kapral2013,LaugaReview,ElgetiReview,Bechinger_RMP2016}. For example, sperm cells have been observed to accumulate at solid walls~\cite{Rotschild1963} and bacteria swim in circles when close to substrates~\cite{Frymier1995} or fluid interfaces~\cite{DiLeonardo2011}. Moreover, synthetic swimmers have been shown to be sensitive to both solid boundaries~\cite{Popescu2009,Howse2015,Uspal2015,Simmchen2016,Schaar2015,Spagnolie2012,Chinappi2016,Malgaretti2017,Bartolo2017} and liquid interfaces~\cite{Stocco2015,Malgaretti2016,Dominguez2016,Dominguez2016_2,Simmchen2017,Isa2017,Malgaretti2018}.
Concerning synthetic swimmers, the presence of boundaries and interfaces is particularly relevant for self-phoretic colloids. Since these active colloids attain their net displacement by generating local gradients of intensive thermodynamic quantities, such as temperature or the (electro)chemical potential~\cite{Anderson1989a,Golestanian2005,Julicher2009,Poon2013}, their active displacement is sensitive to barriers and interfaces, which affect the profile of the local  temperature and the (electro)chemical potential gradients.
\begin{figure}
    \includegraphics[scale=0.4]{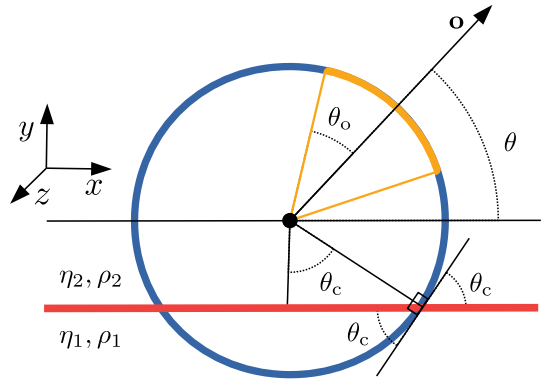}
    \caption{\label{fig:interfaceSystem} Illustration of the system of a self-diffusiophoretic colloid (sphere) at a fluid-fluid interface (red, solid line) between two fluids with dynamic viscosities $\eta_1$ and $\eta_2$ and reactant number densities $\rho_1$ and $\rho_2$. The blue line represents the inert region of the colloid surface, while the orange, solid line represents the catalytic region with opening angle $\theta_{\mathrm{o}}$. The colloid is partially wetted by both fluids and its position with respect to the interface is captured by the contact angle $\theta_{\mathrm{c}}$. The orientation of the colloid with respect to the interface is characterized by the angle $\theta$ between the axis $\bm{o}$ of the colloid and a line parallel to the interface (black, solid line). Here the orientation of the particle corresponds to $\theta_{\mathrm{o}}>0$.}
\end{figure}

In this contribution we analyze the dynamics of self-diffusiophoretic colloids (i.e., particles inducing local gradients in the chemical potentials of certain suspended species) which are adsorbed at a fluid--fluid interface. 
The fluid flow induced by the local stresses is caused by the chemical reaction on the surface of the particle and it will be affected by the presence of two, phase separated fluid phases.
The description of such a system via the standard  coarse-grained approach, within which the relative velocity between the particle and the fluid is accounted for by the so-called phoretic slip velocity on the surface of the particle~\cite{Anderson1989a,Golestanian2005,Popescu2009,Julicher2009,Poon2013}, might be insufficient. Indeed, the phoretic slip velocity has been invoked in those cases in which the imbalance in the local chemical potential is confined to a thin shell around the particle where the Stokes equation is solved analytically~\cite{Anderson1989a}. This approach becomes more complicated if the active colloid is adsorbed at a fluid interface in the presence of a three-phase contact line. 
In this context, we present a novel approach, based on numerical simulations, in which the motion of the self-diffusiophoretic colloid is obtained by using the lattice Boltzmann method in order to construct approximate solutions of the Navier-Stokes equation directly. In our scheme this hydrodynamics solver is combined with an advection and diffusion equation for the reactants.
Such an approach allows us to discuss the reliability of the slip-velocity approach by comparison with previous approximate analytical results~\cite{Malgaretti2016}.
In particular, in the present study we focus on the case in which the two fluids have the same viscosity, for which the approximate analytical model~\cite{Malgaretti2016} predicts the absence of any torque on the particle. Interestingly, our results show that, even in this case, self-phoretic colloids trapped at fluid interfaces reorient their symmetry axis. This reorientation occurs whenever the axis of symmetry of the particle is not perpendicular or parallel to the interface. Indeed, for these cases the presence of the interface affects the velocity profile and leads to net torques on the particle.

The presentation of our study is organized as follows. In Sec.~\ref{sec:latticeBoltzmann} we describe our numerical method based on lattice Boltzmann simulations. In Sec.~\ref{sec:particleInterface} we report our results for the dynamics of self-diffusiophoretic colloids trapped at fluid interfaces, and in Sec.~\ref{sec:conclusions} we summarize our main findings.


\section{Numerical Methods\label{sec:latticeBoltzmann}}
Our system is composed of two phase separated fluids (e.g., oil and water) acting as solvents, the reactants, and the products of the chemical reaction (such as the decomposition of hydrogen peroxide into water and oxygen), and the colloid (see Fig.~\ref{fig:interfaceSystem}). In order to determine the dynamics of the system, we put forward diverse numerical approaches for describing each of these components.

\subsection{The lattice Boltzmann method}
In order to solve the dynamics of the fluids we use the lattice Boltzmann method  (LBM) as it is implemented in the LB3D package~\cite{Benzi1992,Harting2005}.
Within the LBM the fluid phases are described by their discretized single particle distribution functions $f_i^{\sigma}(\bm{r}, t)$, which give the probability of finding a fluid particle of component $\sigma$ at position $\bm{r}$ with velocity $\bm{c}_i$. Our system consists of two species, such as $\sigma_1=\,$oil and $\sigma_2=\,$water, forming two fully segregated phases.
Here, we use a so-called D3Q19 lattice with $19$ discrete velocities $\bm{c}_{i}\,,i=1,...,19$, in three dimensions. We measure times in units of the integration time step $\Delta t$ and lengths in units of the lattice spacing $\Delta x$. These microscopic auxiliary quantities have no physical meaning and their values are chosen to be smaller than any other physically relevant length or time scale. 
Eventually, it turns out to be convenient to fix the magnitudes of $\Delta t$ and $\Delta x$ to unity and to measure times and lengths in units of $\Delta t$ and $\Delta x$, respectively. Accordingly, the actual dimensional values of $\Delta t$ and $\Delta x$ in actual units follow from the smallest length and time scale \textit{with physical meaning}\footnote{For example, in the case of a single colloid suspended in a Newtonian fluid  the smallest relevant length scale is the  size of the colloid and the relevant time scale is its mobility.}.  
The particle distribution functions $f_i^{\sigma}(\bm{r}, t)$ evolve in time due to advection from the neighboring lattice sites and due to collisions among particles at the same lattice site. 
In the following we use the so-called Bhatnagar-Gross-Krook (BGK) collision operator~\cite{Bhatnagar1954}. After some algebra involving the discretization of space, time, and velocities~\cite{Krueger} the time evolution of the distribution  functions $f_i^{\sigma}(\bm{r}, t)$ follows as
\begin{align}
    \label{eq:boltzmannDiscrete} f_i^{\sigma}(\bm{r} + \bm{c}_i \Delta t, t + \Delta t) - f_i^{\sigma}(\bm{r}, t) = -\frac{\Delta t}{\tau^{\sigma}} \left[f_i^{\sigma}(\bm{r}, t) - f_i^{\sigma,eq}(\bm{r} ,t)\right]\, ,
\end{align} 
where the rhs of Eq.~\eqref{eq:eqDistMulti} is the BGK collision operator, $\tau^{\sigma}$ is the relaxation time of the fluid component $\sigma$, and 
$f_i^{\sigma,eq}(\bm{r} ,t)$ is the local equilibrium 
distribution function which, in the small Mach number limit, is given by~\cite{ShanChen1993}
\begin{align}
    \label{eq:eqDistMulti} 
    m^\sigma f_i^{\sigma, eq}(\bm{r},t) = &\, \zeta_i \rho^{\sigma}(\bm{r},t) \biggl(
    1
    + \frac{1}{c_{\mathrm{s}}^2} \bm{c}_i \cdot \bm{\bar{u}}(\bm{r},t) 
    + \frac{1}{2 c_{\mathrm{s}}^4} \left(\bm{c}_i \cdot \bm{\bar{u}}(\bm{r},t)\right)^2 \nonumber\\
    &- \frac{1}{2c_{\mathrm{s}}^2} \bm{\bar{u}}^2(\bm{r},t)
    + \frac{1}{6c_{\mathrm{s}}^6} \left(\bm{c}_i \cdot \bm{\bar{u}}(\bm{r},t)\right)^3
    - \frac{1}{2c_{\mathrm{s}}^4} \bm{\bar{u}}^2(\bm{r},t)\left(\bm{c}_i \cdot \bm{\bar{u}}(\bm{r},t)\right)\biggr),
\end{align}
where $\zeta_i$ are the lattice weights~\cite{Benzi1992,Harting2005} and $m^\sigma$ is the mass of species $\sigma$. The relaxation time is related to the kinematic viscosity of the fluid as $\nu^{\sigma} = (c^\sigma_{\mathrm{s}})^2 \left(\tau^{\sigma} - \frac{{\Delta t}}{2}\right)$, where $c^\sigma_{\mathrm{s}}=\sqrt{\frac{k_BT}{m^\sigma}}$ is the speed of sound in the phase dominated by species $\sigma$. Our numerical approach, in the present form, requires 
all species to have the same mass $m^\sigma=m$, and hence the same speed of sound $c^\sigma_{\mathrm{s}}=c_{\mathrm{s}}$. In the following we choose to fix the lattice time step $\Delta t$ and the lattice spacing $\Delta x$ to unity. In these units, it is common to choose $c_{\mathrm{s}}=\frac{1}{\sqrt{3}}\frac{\Delta x}{\Delta t}$ as the lattice speed of sound in terms of the time step $\Delta t$ and the lattice spacing $\Delta x$~\cite{Krueger}.
Once the distribution functions $f_i^{\sigma,eq}(\bm{r} ,t)$  are  known, it is possible to compute the local mass density of the fluid:
\begin{align}
    \rho^{\sigma}(\bm{r},t) =m \sum_i f_i^{\sigma}(\bm{r},t)\,\,
\end{align}
where $m^\sigma$ is the mass of a single particle of species $\sigma$. The barycentric velocity of the fluid mixture is
\begin{align}
    \bar{\bm{u}}(\bm{r},t) = \sum_{\sigma} \frac{\rho^{\sigma}(\bm{r},t) \bm{u}^{\sigma}(\bm{r},t)}{\tau^{\sigma}} \bigg/ \sum_{\sigma} \frac{\rho^{\sigma}(\bm{r},t)}{\tau^{\sigma}}\,.
    \label{eq:weighedU} 
\end{align} 
Finally, the velocities of the individual fluid components are given by
\begin{align}
\bm{u}^\sigma(\bm{r},t)&=\frac{m}{\rho^\sigma(\bm{r},t)}\sum_i f^\sigma_i(\bm{r},t)\bm{c}_i.
\end{align}

In order to account for multiple solvent phases we follow the method introduced by Shan and Chen~\cite{ShanChen1993}. Within this method the interaction force density acting among distinct species has the form
\begin{align}
    \bm{F}^{\sigma}(\bm{r}, t) = -\psi^{\sigma}(\bm{r}, t) \sum_{\sigma'} \sum_i g_{\sigma \sigma'} \psi^{\sigma'}(\bm{r} + \bm{c}_i\Delta t, t) \bm{c}_i\frac{\Delta t}{\Delta x},
    \label{eq:interaction} 
\end{align} 
where $g_{\sigma \sigma'}$ denotes the interaction strength between the  components $\sigma$ and $\sigma'$; $\psi^{\sigma}$ is a dimensionless pseudo-potential, which is a functional of the mass density. Here, the functional form of $\psi^\sigma$ is chosen as\footnote{The masses of the reactants, the reaction product, and  the colloid enter into the description via their respective equation of motion.}
\begin{align}
    \psi^{\sigma}(\bm{r}, t) = 1-\exp\left[-\frac{\rho^\sigma(\bm{r}, t)}{\rho^\sigma_0}\right]\,.
    \label{eq:psiFunction} 
\end{align} 
where $\rho^\sigma_0$ is a reference density which is related to the bulk properties of the phase dominated by species $\sigma$.
The force density in Eq.~\eqref{eq:interaction} is applied to the fluid by adding a shift to $\bar{\bm{u}}$: 
\begin{align}
    \bar{\bm{u}}'(\bm{r},t) = \bar{\bm{u}}(\bm{r},t) + \frac{\tau^{\sigma} \bm{F}^{\sigma}(\bm{r},t)}{\rho^{\sigma}(\bm{r},t)}\,.
    \label{eq:weighedVelocityChange} 
\end{align} 
Accordingly, in the expression for the equilibrium distribution functions $f_i^{\sigma, eq}(\bm{r},t)$ , $\bar{\bm{u}}$ is replaced by $\bar{\bm{u}}'$ (see Eq.~\eqref{eq:weighedVelocityChange}). 
Values of the interaction strengths  $g_{\sigma \sigma'}$ between distinct species, i.e., $\sigma \neq \sigma'$, which exceed a threshold,  eventually lead to their separation, whereas values of $g_{\sigma \sigma'}$ for $\sigma = \sigma'$ exceeding the threshold give rise to the separation of the liquid and the vapor phases of a certain species~\cite{ShanChen1993}. 
In the following we take $g_{\sigma \sigma'}=0.1$ and $g_{\sigma \sigma}=0$ which leads to an interface with a thickness of ca. $5$ lattice units and to a surface tension of the order of $0.1$ in lattice units.

In order to study the behavior of a colloid suspended at a fluid-fluid interface, a scheme is needed for treating objects, which are large compared with the particles  forming the fluids.
The separation of length scales between the mesoscopic colloidal size and the molecular size of the fluid particles allows one to keep the coarse-grained description for the fluid (via LBM) while  simultaneously treating the colloid as a spherical object characterized completely by its size, position, orientation, and its linear and angular velocities. 
The interaction between the colloid and the fluid gives rise to forces and torques acting on both the colloid and the fluid. 
The technical details of the implementation of the coupling between the colloid and the fluid within LBM are discussed in  Refs.~\cite{Ladd1994_1,Ladd1994_2,Aidun1998,Jansen2011,Frijters2012}. 
In the following we shall outline only the basic features of this method.

The colloid occupies those lattice cells which are inside the spherical colloid (see Fig.~\ref{fig:discreteSystem}).  
As the colloid moves, the configuration of lattice cells occupied by the colloid is updated.
Solid impenetrability is accounted for by bouncing back those contributions  to the fluid flow which attempt to invade the solid boundaries~\cite{Ladd2001,Jansen2011}. 
This is implemented by updating the distribution function after the streaming step according to
\begin{align}
f_i^\sigma(\mathbf{r},t+\Delta t)=f_i^\sigma(\mathbf{r}-\mathbf{c}_i\Delta t,t)
    \label{eq:postcollision-1} 
\end{align}
for all $i$, if the lattice site  $\mathbf{r}-\mathbf{c}_i\Delta t$ is occupied by fluid particles, whereas  if   $\mathbf{r}-\mathbf{c}_i\Delta t$ is occupied by the colloid, the fluid particles are bounced back, i.e., their velocity  is flipped:
\begin{align}
f_i^\sigma(\mathbf{r},t+\Delta t)=f_{i'}^\sigma(\mathbf{r},t),
    \label{eq:postcollision-2} 
\end{align}
where $i'$ is defined as the index corresponding to $\mathbf{c}_{i'}=-\mathbf{c}_{i}$.
This procedure leads to a no-slip boundary condition at the surface of the colloid and to a momentum transfer between the fluid species $\sigma$ and the colloid, which induces a local force density
\begin{equation}
    \mathbf{F}^\sigma(\mathbf{r},t)=\frac{2}{\Delta t} \rho^\sigma(\mathbf{r},t)\mathbf{c}_i
\end{equation}
and a torque density
\begin{equation}
    \bm{T}^\sigma(\mathbf{r},t)=\mathbf{r}(t)\times \bm{F}^\sigma(\mathbf{r},t),
\end{equation}
where $\mathbf{r}(t)$ is the vector pointing from the center of the colloid to the site where bounce-back occurs. 
In order to be consistent the above mentioned bounce-back rule (see Eq.~\eqref{eq:postcollision-2}) has to be modified by accounting for the motion of the colloid~\cite{Ladd2001,Jansen2011}. In order to do so, a correction is  added to the fluid distribution functions~\cite{Ladd2001,Jansen2011}:
\begin{align}
    f_{i}^{\sigma}(\bm{r}, t+\Delta t) = f_{i^{\prime}}^{\sigma}(\bm{r}, t)  - \frac{1}{6} \rho^{\sigma}(\bm{r}, t)\bm{u}_\text{surf}(\bm{r},t)\mathbf{c}_{i'}\frac{(\Delta t)^2}{(\Delta x)^2}\frac{1}{m^\sigma},
    \label{eq:postcollision-3} 
\end{align} 
where $\bm{u}_\text{surf}$ is the local velocity at the surface of the colloid, $\bm{r}$ is the position of the fluid lattice site. (Note that Eq.~\eqref{eq:postcollision-3} holds explicitly in the limit $\Delta t\rightarrow 0$ due to Eq.~\eqref{eq:postcollision-2}.) Consequently, the force density acting on the colloid is modified, too:
\begin{align}
    \mathbf{F}^\sigma(\bm{r}, t) =\frac{1}{\Delta t}\left( 2\rho^{\sigma}(\bm{r}, t)  - \frac{1}{6} \rho^{\sigma}(\bm{r}, t)\bm{u}_\text{surf}(\bm{r},t)\mathbf{c}_{i'}\frac{(\Delta t)^2}{(\Delta x)^2}\right)\mathbf{c}_{i'}\,.
    \label{eq:postcollision-4} 
\end{align} 
Finally, when the colloid moves it occupies pristine cells at its front and it releases cells at its back. In the case of newly occupied cells, the fluid located therein is deleted and its momentum is transferred to the colloid  by adding
\begin{align}
    \mathbf{F}^\sigma(\mathbf{r},t)=-\frac{1}{\Delta t}\rho^\sigma(\mathbf{r},t)\mathbf{u}^\sigma(\mathbf{r},t)
\end{align}
to the rhs of Eq.~\eqref{eq:postcollision-4}.
In the case of cells being released by the colloid, fresh fluid is created at the corresponding sites with velocity  $\bm{u}_\text{surf}(\mathbf{r},t)$ and  density  $\bar{\rho}^\sigma$, where $\bar{\rho}^\sigma$ is obtained by averaging the fluid composition in the direct neighborhood ~\cite{Jansen2011}:
\begin{align}\label{eq:avgden}
    \bar{\rho}^\sigma(\bm{r},t)=\frac{1}{N}\sideset{}{'}\sum_{i}\rho^\sigma(\bm{r}+\bm{c}_{i}\Delta t,t),
\end{align}
where the sum $\sideset{}{'}\sum_{i}$ is restricted to those values of $i$ for which $\bm{r}+\bm{c}_i\Delta t$ is a fluid site; $N$ is the number of these sites. In addition, in order to account for the unknown exact density profile in the close vicinity of a colloid, we apply a density correction algorithm as explained in Refs.~\cite{Jansen2011,Frijters2012}.
In order to conserve momentum, a contribution is added to the force density (Eq.~\eqref{eq:postcollision-4}):
\begin{align}
    \mathbf{F}^\sigma=\frac{1}{\Delta t}\bar{\rho}^\sigma\bm{u}_\text{surf}(\mathbf{r},t)\,.
\end{align}

In order to avoid artifacts during the computation of the Shan-Chen forces acting on the colloid and to tune the wetting properties of the colloid, 
the outermost layer of lattice sites occupied by the colloid is filled with a virtual fluid which is only used during the computation of the Shan-Chen forces acting in the direct vicinity of the colloid. Its density is obtained similarly to Eq.~\eqref{eq:avgden}, but can be tuned by adding an offset $\Delta\rho$ to Eq.~\eqref{eq:avgden}, which controls the wettability of the colloid.
Finally, as usual the Shan-Chen forces are used to compute the force acting on the colloid:
\begin{align}
    \bm{F}(t)=\sum_{\bm{r}}\sum_\sigma \bm{F}^\sigma(\bm{r},t)(\Delta x)^3\,.
    \label{eq:cont-ang}
\end{align}
Accordingly, by tuning the magnitude of $\Delta \rho$, it is possible to control the contact angle of the colloid~\cite{Jansen2011,Frijters2012}.

\subsection{Dynamics of the reactant and the reaction product\label{sec:activity}}
\begin{figure}
    \includegraphics[scale=0.3]{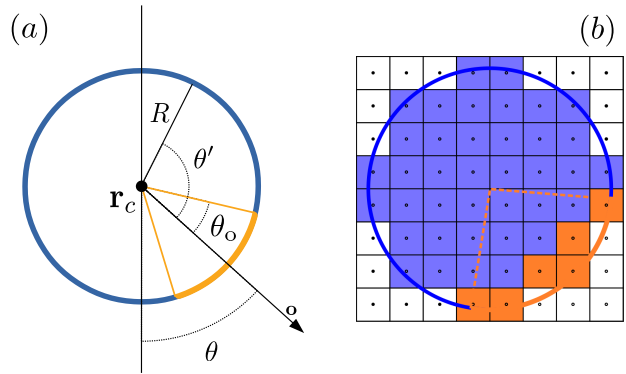}
    \caption{\label{fig:discreteSystem} The solid sphere (a) represents the idealized colloid, while the cubic cells represent the lattice Boltzmann cells (b). Fluid cells are white, colloid cells are blue, and catalytically active cells are orange. The center of the cells is indicated by a black dot.} 
\end{figure}

We assume that the two fluids forming the adjacent phases act as reservoirs of the reactant and of the reaction product such that the mass density $\rho^r$ of the reactant is regarded as to be homogeneous in the bulk of both fluid phases. Close to the interface the mass density of the two fluid phases varies. In the following we assume that the ratios 
\begin{equation}
    \mathcal{C}^\sigma=\frac{f^r(\bm{r})}{f^\sigma(\bm{r})}
\end{equation}
of the number densities of the reactant, $f^r$, and those of the fluid phases,
\begin{equation}
   f^\sigma=\sum_i f_i^\sigma\,,
\end{equation}
are kept constant throughout both fluid phases. Accordingly,  the mass density of reactants at a given position is defined as:
\begin{align}
    \rho^{r}(\bm{r}) = m^{r} \sum_{\sigma} \mathcal{C}^{\sigma} f^{\sigma}(\bm{r})\,,
    \label{eq:reactantsdensity} 
\end{align} 
where the sum runs over all fluid species $\sigma$ and $m^r$ is the mass of a reactant molecule.
Equation~\eqref{eq:reactantsdensity} implies that $\rho^{r}(\bm{r})=\kappa m^r f^r(\bm{r})$ where $\kappa$ is the number of fluid species. 
Concerning the interaction between the reactant molecules and the colloid, we assume that the surface of the colloid is covered by a  catalyst with axial symmetry (see Fig.~\ref{fig:discreteSystem}(a)).
The strength of the chemical reaction is controlled by the surface activity $\xi$, which determines the rate at which reactant molecules are converted into solute molecules.
A simple choice of the surface activity $\xi$ is given by
\begin{align}
    \label{eq:activity}
    \xi(\theta') = \xi_0
    \begin{cases}
        1, &\mathrm{if} \quad \theta' < \theta_\mathrm{o} - \Delta \\
        0, &\mathrm{if} \quad \theta' > \theta_\mathrm{o} + \Delta \\
        \dfrac{\theta_\mathrm{o}-\theta'+\Delta}{2 \Delta}, &\mathrm{otherwise\,.}\\
    \end{cases}
\end{align} 
The prefactor $\xi_0$ is the base activity, $\theta'$ is the azimuthal angle (see Fig.~\ref{fig:discreteSystem}), $\theta_{\mathrm{o}}$ is the opening angle which defines the size of the catalytic cap (see Fig.~\ref{fig:discreteSystem}(a)), and $\Delta$ is an interpolation length.
This corresponds to full activity for angles well within $\theta_{\mathrm{o}}$, zero activity outside, and a linear interpolation in between. Without this linear interpolation (i.e., for $\Delta = 0$), small rotations of the colloid may not change the activity, due to the roughness of the colloid surface\footnote{This "roughness" is due to the discrete nature of the LBM.}.
Optimally, the value of $\Delta$ should correspond to ca. one lattice cell, which is accomplished if $\Delta \approx \Delta x / R$, where $R$ is the radius of the colloid. This is the quantity which we have used in our simulations.

The local flow $\mathbf{J}$ of the reaction product,  produced by the chemical reaction at the surface of the colloid, into a fluid cell at $\bm{r}$ per  time is given by 
\begin{align}
    \label{eq:fluxImplementation}
    \bm{J}(\bm{r})=\alpha
    \rho^{rp}(\bm{r})  \Xi(\bm{r})\mathbf{n}(\bm{r})
\end{align} 
where $\alpha$ is the reaction rate, $\mathbf{n}$ is the local normal at the colloid surface pointing towards the fluid phase, $\rho^{rp}$ is the reaction product mass density, and $\Xi(\bm{r})$ denotes the activity of the colloid cells neighboring the fluid cell at $\bm{r}$:
\begin{align}
    \Xi(\bm{r}) =
    \begin{cases}
    1, & \text{if}\,\,\sum_{\bm{r}_{\mathrm{p}}} \xi(\theta')>0,\\
    0, & \text{otherwise},
    \end{cases}
    \label{eq:totalactivity} 
\end{align}
where the sum is performed only over neighboring cells $\bm{r}_{\mathrm{p}}$ inside  the colloid, and $\Theta$ is the Heaviside function. 
This recipe prevents fluid cells in contact with more than one catalyst cell to be exposed to a larger flux, induced by the discretization of the shape of the colloid.
Equations~(\ref{eq:fluxImplementation}) and (\ref{eq:totalactivity}) enforce a constant flux per area even in the discrete representation of the colloid.

In the following we specialize on the case of a single reaction product (our approach can be easily extended to an arbitrary number of them). Moreover we focus on the small P\'eclet limit, in which the advection contribution to the time evolution of the density of the reaction product is negligibly small as compared with the diffusion contribution. Accordingly, the dynamics of the mass density of the reaction product is decoupled from that of the two fluid phases. The former appears as an additional scalar field defined on the LBM lattice, such that a cell at lattice position $\bm{r}$ contains a mass density $\rho^{rp}(\bm{r})$:
\begin{equation}
    \rho^{rp}(\bm{r})=m^{rp}\sum_\sigma \mathcal{C}_\sigma^{rp}(\bm{r})f^\sigma(\bm{r})
\end{equation}
where, as for the reactant (see Eq.\eqref{eq:reactantsdensity}), we have introduced the number density ratios $\mathcal{C}_\sigma^{rp}(\bm{r})$ and we have accounted for their spatial variation.
In order to  
ensure that $\rho^{rp}(\bm{r})$ reaches a steady state within the simulation box with periodic boundary conditions along the three spatial directions, we introduce a homogeneous sink term with constant decay rate\footnote{This term induces an additional decay length $\Lambda\propto\sqrt\frac{D}{\chi}$. The contribution of the sink term to the velocity of the colloid is disregardable provided that $\Lambda\gg \lambda$ where $\lambda$ is the decay length of the strength of the interaction between the colloid and the reaction product.} $\chi$. Hence the mass density of the reaction product is governed by the partial differential equation
\begin{align}
    \label{eq:diffusionSinks} \frac{\partial}{\partial t}\rho^{rp} &= D \bm{\nabla} \cdot \left( \bm{\nabla} \rho^{rp} + \beta \rho^{rp} \bm{\nabla} U\right) - \chi \rho^{rp}
\end{align} 
with the boundary condition
\begin{align}
    \bm{J}\cdot\mathbf{n}|_{|\bm{r}-\bm{r}_c|=R}&=\alpha \rho^{rp}\xi(|\bm{r}-\bm{r}_c|=R)\,.
\end{align}
Here $|\bm{r}-\bm{r}_c|$ is the distance of a fluid cell from the center of mass of the colloid, located at $\bm{r}_c$ (see Fig.~\ref{fig:discreteSystem}). In Eq.~(\ref{eq:diffusionSinks}) we have introduced the interaction potential between the colloid and the reaction product
\begin{align}
    U(\bm{r}-\bm{r}_c) = 
    \begin{cases}
         (l-|\bm{r}-\bm{r}_c|)F_0\,, & \mathrm{ if } \;\; R < |\bm{r}-\bm{r}_c|<l\,,\\
        0\,, & \mathrm{otherwise}\,,
    \end{cases}
    \label{eq:interactionPotentialForm} 
\end{align} 
where $F_0$ and $l$ are the strength and the range of the potential, respectively. In the following we assume $l=4$ lattice units and $\beta F_0 l=5\times 10^{-5} (\Delta x)^3$ where $\beta$ is the inverse thermal energy. This potential has no angular dependence, i.e., it is the same for the catalytic and the inert side of the colloid surface. Equation~(\ref{eq:diffusionSinks}) is solved via a finite-difference scheme on the same grid as the one used by the LBM.
Finally, the potential $U$, together with the non-equilibrium mass density profile $\rho^{rp}$ of the reaction product, induces a pressure gradient on the fluid phases as~\cite{Anderson1989a}
\begin{align}
    \bm{\nabla} p(\bm{r}) = \dfrac{1}{m^{rp}}\rho^{rp}(\bm{r}) \bm{\nabla} U(\bm{r}).
    \label{eq:pressureGradientSecondTerm} 
\end{align} 

\section{Results\label{sec:particleInterface}}
After validating our numerical scheme against analytical predictions for the velocity of a self-diffusiophoretic colloid in a homogeneous fluid~\cite{Popescu2010} (see Appendix~\ref{sec:homogeneousFluid}),  we study the dynamics of a self-diffusiophoretic colloid adsorbed at a fluid interface.
All simulations are initialized by equilibrating the interface after the colloid has adsorbed; then the chemical reaction at the surface of the colloid is turned on.
In the following, we analyze the dynamics of the colloid in the case that the viscosities of the two fluids are equal, $\eta_{\mathrm{1}} = \eta_{\mathrm{2}}=\eta$, and for various values of the opening angle $\theta_{\mathrm{o}}$ of the catalytic cap and of the contact angle $\theta_{\mathrm{c}}$ (see Fig.~\ref{fig:interfaceSystem}), which characterizes the adsorption of the colloid at the interface. 
Concerning the mass density of the reactant we consider two situations.
In the first one, both fluids have the same mass fraction of reactant molecules: i.e., their mass fractions are assumed to be equal
\begin{align}
    \mathcal{C}^{\sigma_1} = \mathcal{C}^{\sigma_2} \neq 0.
    \label{eq:twoActive}
\end{align} 
In the second case one has
\begin{align}
    \mathcal{C}^{\sigma_1} \neq 0,\,\, \mathcal{C}^{\sigma_2} = 0,
    \label{eq:oneActive}
\end{align} 
so that one fluid does not contain any reactant molecules. 
Finally, we study two different scenarios: one, in which the colloid can move freely along the interface, and another one, in which the lateral colloid position is fixed by an external force.

\subsection{Equal reactant mass fractions\label{sec:twoActive}}
First, we consider the case in which the reactants are suspended with equal mass fraction in both fluid phases (Eq.~\eqref{eq:twoActive}). In this case certain symmetries can be identified, depending on the relative orientation $\theta$  of the axis of the colloid with respect to the plane of the interface: i) The fore-aft symmetry for $\theta = \pm \pi / 2$, i.e., if the axis of the colloid is parallel to the normal of  the interface. (ii) For $c_2 = c_1 \neq 0$ an additional mirror symmetry appears about $\theta = 0$, i.e., if the axis of the colloid lies within the plane of the interface. 

As a first case, we study the dynamics of a colloid which is partially covered by catalyst, i.e., here  $\theta_{\mathrm{o}} = \pi / 4$, and has a contact angle $\theta_{\mathrm{c}} = \pi / 2$. 
\begin{figure}
    \includegraphics[scale=1.1]{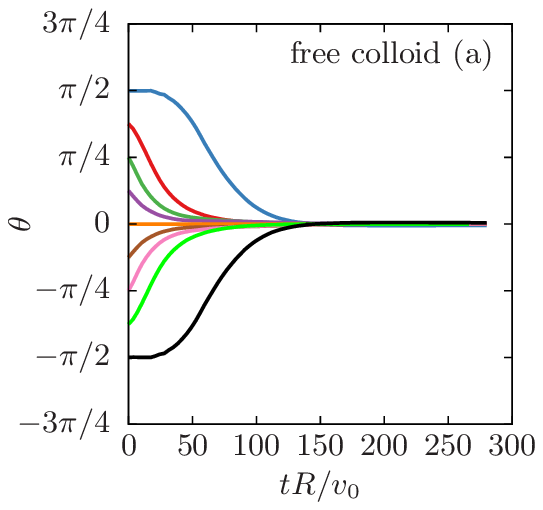}
    \includegraphics[scale=1.1]{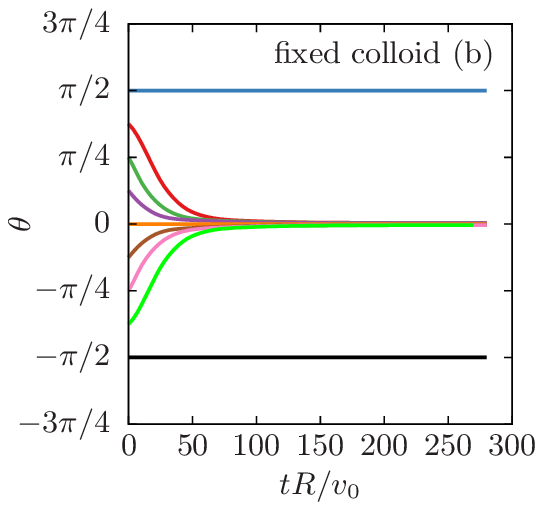}
\caption{
    \label{fig:janusTrajectory}
    Angle $\theta$ of the orientation of the colloid with respect to the interface (see Fig.~\ref{fig:interfaceSystem}) as a function of time, normalized by the time $t_0$ it takes for a half-covered colloid to move a distance which equals its own radius, $t_0=R/v_0$, for the case of a free colloid  (panel (a)) and for the case of a fixed colloid (panel (b)). The velocity $v_0$ is the velocity the very same colloid attains in a homogeneous fluid characterized by the same viscosity (see Appendix~\ref{sec:homogeneousFluid}).
    The lines correspond to various initial orientations $\theta(t=0)$ of the colloid [$\theta(t=0)=0,\pi/8,\pi/4,3\pi/8,\pi/2$] for 
    $\theta_{\mathrm{o}} = \pi / 4$. In the case of the free colloid, the trajectories have been smoothed by integrating over a shifting time-window $10^4\Delta t$ wide. The lines are symmetric with respect to $\theta=0$.}
\end{figure}
Figure~\ref{fig:janusTrajectory} shows the orientation $\theta(t)$ of the colloid as a function of time for a free colloid (Fig.~\ref{fig:janusTrajectory}(a)) as well as for a colloid the center of mass of which is kept at a fixed position by an external force (Fig.~\ref{fig:janusTrajectory}(b)). 
Interestingly, the dynamics in the two setups are quite similar, in that  the catalytic reaction induces a net torque on the colloid which leads $\theta$ to approach a steady state with  $\theta(t=\infty)\,\equiv\theta_\infty=0$. 
We note that, while for the moving colloid $\theta_\infty\,=0$ is the only steady state, for the fixed colloid $\theta_\infty=\pi/2$ is also a steady state. This difference is due to the fact that if the colloid is not fixed, its center of mass will move also along the direction normal to the interface. Hence, the distinction between the two cases emphasizes that $\theta_\infty=\pi/2$ is not stable with respect to fluctuations of the position of the center of mass of the colloid.

\begin{figure}
    \includegraphics[scale=1.1]{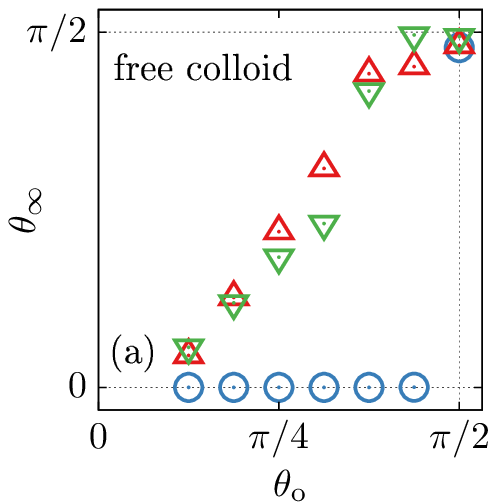}
    \includegraphics[scale=1.1]{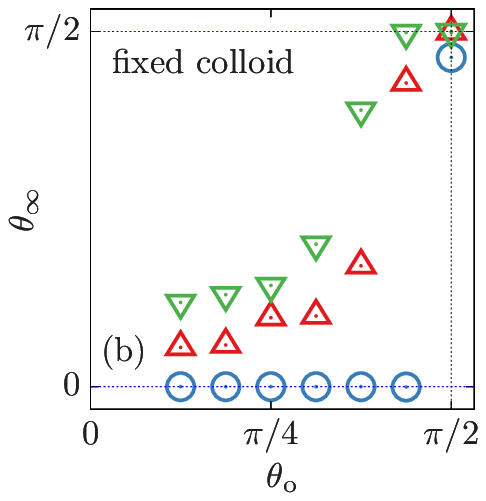}
\caption{
    \label{fig:steadyStates}
    Steady state angle $\theta(t=\infty)\,\equiv\theta_\infty$ as a function of the opening angle $\theta_{\mathrm{o}}$.
    Blue, red, and green symbols correspond to $\theta_\mathrm{c}=\pi/2,\,0.55\pi$, and $0.6\pi$, respectively. Panels (a) and (b) correspond to a free and a fixed colloid, respectively. 
    The data are symmetric with respect to $\theta_\infty=0$ (not shown).
}
\end{figure}

Next, we study the dependence of the dynamics of the colloid on the areal size of the catalytic cap, characterized by $\theta_{\mathrm{o}}$, and on the contact angle $\theta_\mathrm{c}$.
Figure~\ref{fig:steadyStates} shows the stable steady state orientations $\theta_\infty$ as a function of the opening angle $\theta_{\mathrm{o}}$ for various contact angles of a free colloid (Fig.~\ref{fig:steadyStates}(a)) and of a colloid the center of mass of which is kept fixed (Fig.~\ref{fig:steadyStates}(b)).
Interestingly, both panels of  Fig.~\ref{fig:steadyStates} show that for opening angles $\theta_{\mathrm{o}} \neq \pi / 2$ the steady state orientation of the colloid is $\theta_\infty\neq \pi/2$, i.e., the colloid attains a steady translation along  the interface, because the driving force points into the axial direction of the colloid and thus provides a lateral component.
In particular, $\theta_\infty$ grows upon increasing $\theta_\mathrm{o}$ for both  $\theta_\mathrm{c}=0.55\pi$ (red  upward triangles)  and $\theta_\mathrm{c}=0.6\pi$ (green  downward triangles).
Finally, for $\theta_{\mathrm{o}} = \pi / 2$, the stable steady state is $\theta_\infty = \pm \pi / 2$ for all contact angles $\theta_{\mathrm{c}}$
which leads to a vanishing velocity along the interface. 
We note that the case of contact angle $\theta_{\mathrm{c}} = \pi / 2$ (blue circles) is peculiar because in this case the steady state orientation is $\theta_\infty=0$ which implies maximum velocity along the interface. This result holds for all opening angles $\theta_{\mathrm{o}} < \pi/2$. For $\theta_{\mathrm{o}} = \pi/2$ with $\theta_{\mathrm{c}} = \pi/2$ we face numerical difficulties due to the symmetry of the problem. In fact, for $\theta_{\mathrm{c}} = \pi/2$ the center of mass of the colloid lies at the interface and if the colloid is half-covered ($\theta_{\mathrm{o}} = \pi/2$) the only non-motile state requires $\theta=0$. However, when $\theta\rightarrow 0$, the typical size of the active site being wetted by one of the two fluid phases becomes comparable to the lattice constant and, for $\theta_{\mathrm{c}} = \pi/2$ and $\theta_{\mathrm{o}} = \pi/2$, our discrete numerical approach would require larger ratios of the colloid size and the lattice constant which, however, we could not explore.

\begin{figure}
    \includegraphics[scale=1.1]{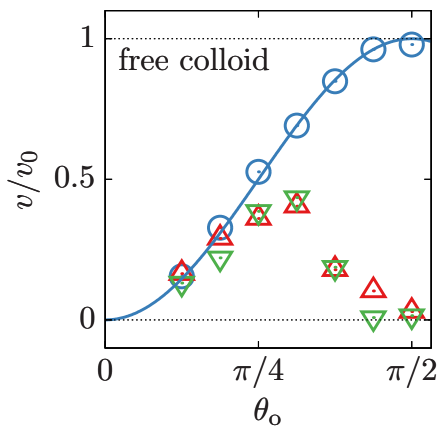}
    \caption{
    \label{fig:velsteadyStates}
    Lateral velocity $v$ along the interface, normalized by the velocity $v_0=v(\theta_\mathrm{o}=\pi/2)$ ($v_0= 0.00175 $ in LBM units), of a free Janus colloid ($\theta_{\rm{o}}=\pi/2$) in a homogeneous fluid  as a function of the opening angle $\theta_{\mathrm{o}}$ for various contact angles:
    blue, red, and green symbols correspond to $\theta_{\mathrm{c}}=\pi/2\,,0.55\pi$, and $0.6\pi$, respectively.
    The blue line represents the velocity of a free Janus colloid in a homogeneous bulk fluid as function of $\theta_{\mathrm{o}}$.
}
\end{figure}

In the case in which the colloid can move freely, at steady state its lateral velocity along the interface depends on both the contact angle and the opening angle. Figure~\ref{fig:velsteadyStates} shows that for $\theta_{\rm c}=\pi/2$ the maximum speed is obtained for $\theta_{\rm{o}}=\pi/2$ and it equals the one ($v_0$) obtained in a homogeneous fluid. This tells  that, under the condition of equal viscosity among the two fluid phases, due to the symmetry of the problem, the interface does not affect the fluid flow and hence the motion of the colloid. In contrast, Fig.~\ref{fig:velsteadyStates} shows that, for $\theta_{\rm c}\neq \pi/2$, $v$ has a non-monotonous dependence on the opening angle. Indeed, $v$ vanishes for $\theta_{\rm{o}}=0$ (i.e., for a passive colloid) and for $\theta_{\rm{o}}=\pi/2$, and it attains a maximum for $\theta_{\rm{o}}\simeq \pi/4$. This non-monotonous dependence of $v$ on $\theta_{\rm o}$ can be understood by relating the data in Fig.~\ref{fig:velsteadyStates} to those shown in Fig.~\ref{fig:steadyStates}(a). Indeed, for $\theta_{\rm{o}}\rightarrow 0$ at steady state the axis of the colloid lies in the plane of the interface (i.e.,  $\theta(t=\infty)\rightarrow 0$), whereas upon increasing $\theta_{\rm{o}}$ the steady state orientation, $\theta_\infty$, also grows and so does the steady state velocity. For even larger values of $\theta_{\rm{o}}$ the value of $\theta_\infty$  increases until, for $\theta_{\rm {o}}\rightarrow \pi/2$, one has  $\theta_\infty \rightarrow \pi/2$ which implies $v\rightarrow 0$.

The rotation of the symmetry axis of the colloid we have just described is rather counter-intuitive, in particular for $\theta_\mathrm{c}=\pi/2$ and $\eta_1=\eta_2$. 
Indeed, the equality of the viscosity of the two fluids and the lack of accumulation of the reaction product at the interface enforce the symmetry of the mass density profile of the reaction product about the symmetry axis of the colloid and therefore of the local pressure gradients acting on the fluid. Hence, at first glance, in this scenario one might expect no net torque. 
Actually, for this case a simplified analytical model~\cite{Malgaretti2016}, which disregards the fluid flow in the vicinity of the three-phase contact line, predicts that there is no reorientation of the colloid  at all.
In the following, we show that the observed torque is due to the fact that the mobility is not homogeneous along the surface of the colloid. Indeed, the boundary conditions imposed on the fluid velocity by the presence of the interface lead to an effective local mobility the spatial variation of which generates the torque. 
In order to highlight this relationship, we determine the net velocity, which a passive colloid attains if it is  pushed by an external local force density  $f_{\bm{r}_0}(\bm{r})$,  localized at $\mathbf{r}_0$ near the colloid surface  and acting towards the center of the colloid:
\begin{align}
    \label{eq:deltaForce} 
    f_{\bm{r}_0}(\bm{r}) = F \frac{\bm{r}_{\mathrm{c}} - \bm{r}}{\left|\bm{r}_{\mathrm{c}} - \bm{r}\right|} \delta \left(\bm{r} - \bm{r}_0\right)\,,
\end{align} 
where $\delta \left(\bm{r} - \bm{r}_0\right)$ is the Dirac delta function, $F$ is a force, and $\bm{r}_\mathrm{c}$ is the position of the center of mass of the colloid. 
In these corresponding simulations, the colloid attains a steady state in which it rotates with constant angular velocity $\boldsymbol{\omega}$. We are interested only in its component $\omega_{z}(t=\infty) = \bm{e}_{\mathrm{z}} \cdot \bm{\omega}(t=\infty)$ along the $z$-direction\footnote{Note that in general, the mobility coefficient is a tensor because both the applied force and the angular velocity are vectors. However, for the self-phoretic colloid the local forces are always perpendicular to the colloid surface, and we are  interested only in the $z$-component of the angular velocity. For this purpose the description of Eq.~\eqref{eq:mobilityCoefficient} suffices.} .
We introduce the dimensionless inverse mobility coefficient
\begin{equation}
    \gamma(\bm{r}_0)\, :=\,\gamma_0 \frac{R\omega(t=\infty)}{F}\,,
    \label{eq:mobilityCoefficient} 
\end{equation} 
as the ratio of the angular velocity and the magnitude of the applied force\footnote{Since the force is acting radially, the torque on the colloid is zero.}, where $R$ is the radius of the colloid, $\gamma_0=6\pi \eta R$ is the friction coefficient of the colloid; $\gamma(\bm{r}_0)$ depends on $\bm{r}_0$ via $\omega(t=\infty)$.

\begin{figure}
    \includegraphics[scale=1.1]{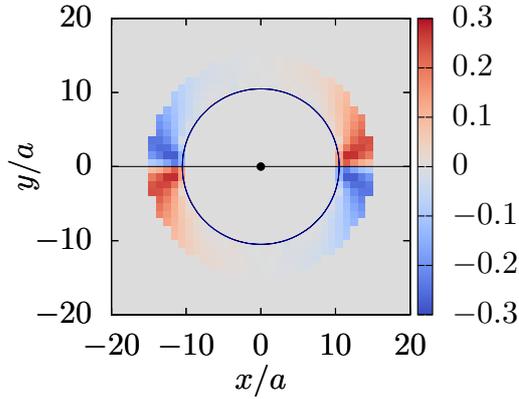}
\caption{
    \label{fig:deltaForce}
    Inverse mobility coefficient (red/blue color) $\gamma$ (Eq.~\eqref{eq:mobilityCoefficient}) as a function of the position in the $x$-$y$ plane through the colloid center (black dot). The interface lies in the $x$-$z$ plane located at $y=0$ (black horizontal line).
    The blue color tells that a force at this position leads to a negative angular velocity $\omega_z<0$, whereas a force in cells with red color leads to a positive angular velocity $\omega_z>0$.
    The inverse mobility coefficient is only shown for positions which are at most $l = 4$ lattice units away from the colloid ($R< r < R+l$), where $l$ is the range of the interaction potential (see Eq.~\eqref{eq:interactionPotentialForm}).
    The blue circle represents the colloid.
}
\end{figure}

\begin{figure}[t]
    \includegraphics[scale=0.9]{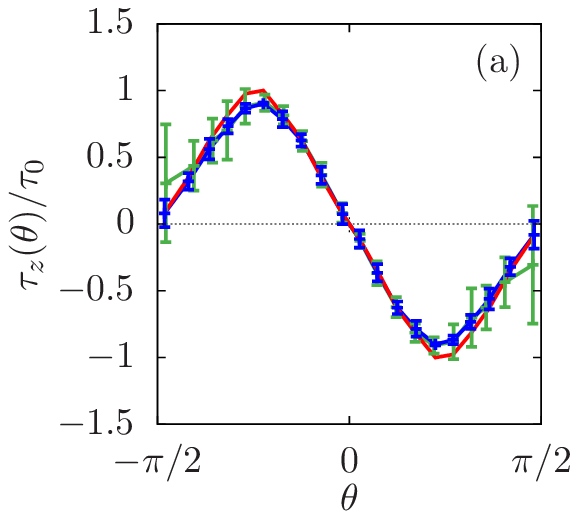}
    \includegraphics[scale=0.9]{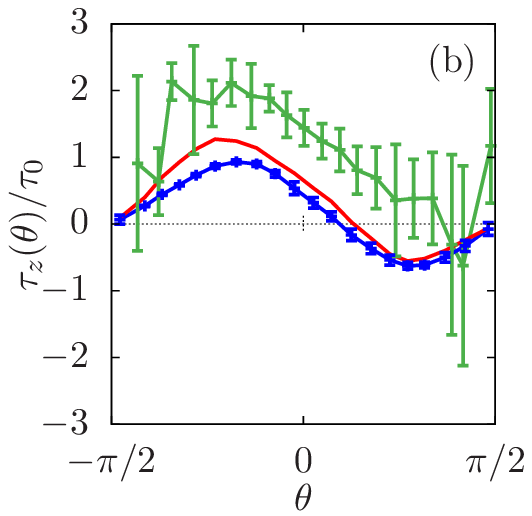}
    \includegraphics[scale=0.9]{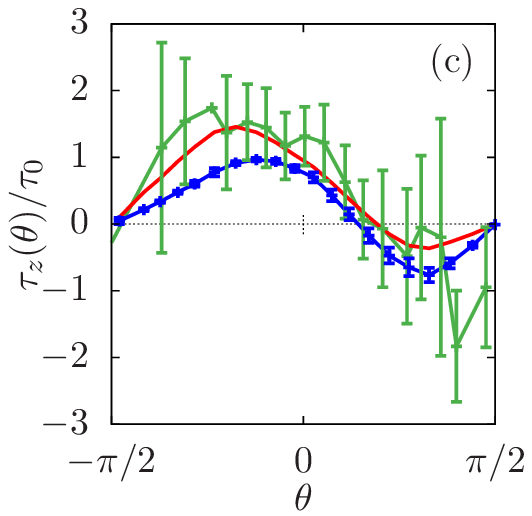}
\caption{
    \label{fig:pi4torque}
    Component $\tau_z$ in the $z$-direction of the torque on the colloid as a function of the rotation $\theta$.
    Blue (green) solid lines correspond to the LBM measured torque for a fixed (free) colloid; the error bars indicate the standard error.  (Note that for certain values of $\theta$ the blue, green, and red lines totally overlap and thus are not visible.) 
    Red solid lines correspond to the torque calculated via Eq.~\eqref{eq:deltaForceIntegral2D} for
    $\theta_{\mathrm{o}} = \pi / 4$
    with (a) $\theta_{\mathrm{c}} = 0.5 \pi$, 
     (b) $\theta_{\mathrm{c}} \approx 0.55 \pi$, and (c) $\theta_{\mathrm{c}} \approx 0.6 \pi$.  The torque components are normalized by $\tau_0\equiv\tau(\theta=-\pi/4,\theta_{\rm{c}}=\pi/2)$ as obtained from Eq.~\eqref{eq:deltaForceIntegral2D} with $\theta_\mathrm{c}=\pi/2$ ($\tau_0=0.04$ in LBM units). 
     The error bars are due to the discretization of the colloid surface (for both fixed and mobile colloids) and due to the motion normal to the interface (mobile colloid, green lines).
}
\end{figure}

In Fig.~\ref{fig:deltaForce}, the dimensionless inverse mobility coefficient is shown as a function of the $x$ and $y$ positions in a plane perpendicular to the interface (black horizontal line in Fig.~\ref{fig:deltaForce}) and passing through the colloid center of mass. This highlights the point symmetry of the dimensionless inverse mobility and that forces closer to the interface give rise to larger torques and hence higher values of the steady angular velocity $\omega_{z}$. 
In the small Reynolds number regime, which is valid for the motion of the present diffusiophoretic colloid, the Navier-Stokes equation reduces to the (linear) Stokes equation. Hence, the local mobility allows us to determine the torque on the colloid in $z$-direction for any given force density $\bm{f}(r)$, as long as the force acts on the colloid along the radial direction. 
For the case of a self-diffusiophoretic colloid, the force density distribution, due to \textit{ph}oresis, is given by $f_{ph}(\bm{r}) = \frac{\partial U(r)}{\partial r} \frac{\rho^{rp}(\bm{r})}{m^{rp}}$ (see Eq.~\eqref{eq:pressureGradientSecondTerm}) where here $r$ indicates the distance from the center of mass of the colloid.
Accordingly, the local mobility allows us to determine the torque $\tau_z=\boldsymbol{\tau}\cdot\mathbf{e}_z$ exerted on the colloid by the density profile of the reaction product:
\begin{align}
    \label{eq:deltaForceIntegral2D} 
     \tau_{\mathrm{z}} = B L \int_{-L}^L \int_{-L}^L 
     \mathrm{d}x\, \mathrm{d}y f_{ph}(x, y, 0) \gamma(x, y, 0),
\end{align} 
where $B$ is a dimensional fitting parameter, and $L$ is the size of the simulation box.
Actually, Eq.~\eqref{eq:deltaForceIntegral2D}  approximates the total torque in that we are considering solely the contributions to the torque stemming from the plane perpendicular to the interface and passing through the center of mass of the colloid (see Fig.~\ref{fig:deltaForce}).
Contributions stemming from the rest of the surface of the colloid are accounted for by the fitting parameter $B$. We emphasize that the same value of the parameter $B$, i.e., $B=5.2\,\Delta x$, fits well the data for  colloids with various opening angles $\theta_\mathrm{o}$ as shown in Figs.~\ref{fig:pi4torque}, \ref{fig:pi2torque}, \ref{fig:3pi8torque}, and \ref{fig:pi8torque}. The validity of this approach will be discussed \textit{a posteriori}, i.e., by comparing it with the values extracted from the corresponding lattice Boltzmann simulations.

Figure~\ref{fig:pi4torque} provides a comparison of the torque calculated via Eq.~(\ref{eq:deltaForceIntegral2D}) with the torque obtained directly from the lattice Boltzmann simulations for a colloid with opening angle $\theta_{\mathrm{o}} = \pi / 4$ and for  three contact angles. (For a discussion of these kind of results for various opening angles see Appendix~\ref{sec:additional}.)
Interestingly, Fig.~\ref{fig:pi4torque}(a) shows that for a contact angle $\theta_{\mathrm{c}} = \pi / 2$ the prediction of Eq.~\eqref{eq:deltaForceIntegral2D} agrees very well with the torque obtained from the simulations for both a free and a fixed colloid. As expected, Fig.~\ref{fig:pi4torque}(a) shows that the torque on the colloid is zero for the steady states characterized by  $\theta=0,\pm\frac{\pi}{2}$.  

For contact angles $\theta_\text{c}\neq\pi / 2$ (see Figs.~\ref{fig:pi4torque}(b) and \ref{fig:pi4torque}(c)), the torques on the free and on the fixed colloid are no longer equal. 
This is expected, because for these contact angles the lateral motion along the interface leads to an additional contribution to the torque~\cite{Pozrikidis2007}.
This explains the mismatch between the prediction of Eq.~\eqref{eq:deltaForceIntegral2D} and the torque of the free colloid obtained from the simulations shown in Figs.~\ref{fig:pi4torque}(b) and  \ref{fig:pi4torque}(c).
However, this additional torque is absent for the fixed colloid. This explains the good agreement between the prediction of Eq.~\eqref{eq:deltaForceIntegral2D} and the torque on the fixed colloid calculated from the simulations shown in Figs.~\ref{fig:pi4torque}(b) and  \ref{fig:pi4torque}(c).

\subsubsection{Inhomogeneous reactant densities\label{sec:oneActive}}
Up to now, we have studied the case in which both fluid phases contain reactant molecules.
Here, we consider the system in which one of the two fluids does not contain the reactant (Eq.~\eqref{eq:oneActive}), yet the product of the chemical reaction can diffuse, with equal diffusivity, in both fluid phases.
Clearly, in this case the point symmetry of $\tau_z(\theta)$ at $\theta = 0$ is broken even for the contact angle $\theta_{\mathrm{c}} = \pi / 2$, whereas its mirror-symmetry at $\theta_{\mathrm{c}} = \pm \pi / 2$ remains. 

Since we have not observed major discrepancies between the case of a moving colloid and that of a colloid fixed in space, in the following we focus on the case of a fixed colloid, because it is computationally easier.
If the catalytic cap is fully immersed in that  fluid phase which  contains the reactant, the dynamics of the colloid is identical to that observed if the reactant is dissolved in both fluid phases. From geometric considerations (see Fig.~\ref{fig:interfaceSystem}), it follows that the catalytic cap is exposed to the fluid without reactant if $\theta\geq\theta_1$ with
\begin{align}
    \theta_1 = (\theta_{\mathrm{c}} - \pi / 2) - \theta_{\mathrm{o}},
    \label{eq:thresholdAngle1} 
\end{align} 
and it is fully submerged in the fluid without reactants for $\theta\geq\theta_2$ with
\begin{align}
    \theta_2 = (\theta_{\mathrm{c}} - \pi / 2) + \theta_{\mathrm{o}}.
    \label{eq:thresholdAngle2} 
\end{align} 
Accordingly, as long as $\theta < \theta_1$ the catalytic cap does not get into contact with the fluid without reactant and hence 
the colloid 
will behave exactly as in the case in which both fluids contain the reactant. If $\theta>\theta_1$, a large gradient of the solute concentration occurs close to the interface due to the lack of reaction in the other fluid phase. This drives the rotation of the colloid further towards the fluid without reactant, until the catalytic cap becomes fully submerged in this fluid, i.e., $\theta\,\geq\,\theta_2$. At this point the system becomes "passive" because there is no longer production of solute and thus the colloid does not move anymore.

\begin{figure}
    \includegraphics[scale=1.1]{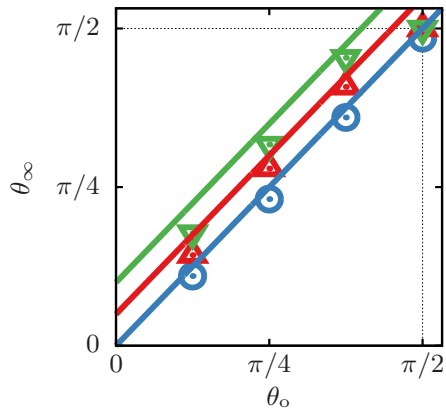}
\caption{
    \label{fig:steadyStatesInactive}
    Steady state angle $\theta_\infty$ as a function of the opening angle $\theta_{\mathrm{o}}$. Symbols are simulation data; blue circles: $\theta_{\mathrm{c}} = \pi / 2$; 
    red upward triangles: $\theta_{\mathrm{c}} \approx 0.55 \pi$; 
    green downward triangles: $\theta_{\mathrm{c}} \approx 0.6 \pi$. 
    The blue, red, and green lines show the threshold angle $\theta_2$ (Eq.~\eqref{eq:thresholdAngle2}) for the corresponding contact angle $\theta_\mathrm{c}$ as a function of the opening angle $\theta_\mathrm{o}$.
}
\end{figure}

Figure~\ref{fig:steadyStatesInactive} shows the steady states $\theta_\infty$ of the colloid as a function of the opening angle. We remark that Fig.~\ref{fig:steadyStatesInactive} does not show all stable steady states with $\theta > \theta_2$, i.e., when the system becomes passive due to the lack of chemical reactions. 
Interestingly, the steady states shown in Figure~\ref{fig:steadyStatesInactive} are in good agreement with our simple geometrical estimate (Eq.~\eqref{eq:thresholdAngle2}) except for a constant offset.
We speculate that this offset is a numerical artifact because a small concentration of reactant molecules can occur in the interface region. (We recall that in the lattice Boltzmann scheme the fluid-fluid interface has a finite thickness, which in the present simulations is ca.  $5$ lattice constants.)
In particular, within the Shan-Chen model the density of the fluid, which contains the reactant, is not zero everywhere because the two fluids do not completely demix.
Therefore, when the colloid protrudes far into the fluid without reactant, there is a torque rotating it back towards the fluid with reactant, in line with the results from Section~\ref{sec:twoActive}.
Finally, we note that for a  Janus colloid, i.e.,  $\theta_\mathrm{o}=\pi/2$, and with $\theta_\mathrm{c}>\pi/2$ the steady state $\theta_\infty = \pi / 2$ is independent of the contact angle, because the Janus colloid is always in contact with the fluid with reactant.

\section{Conclusions\label{sec:conclusions}}

We have presented a novel numerical approach which is capable of capturing the dynamics of self-phoretic colloids even in the presence of a fluid interface which affects the boundary condition of the fluid velocity close to the surface of the colloid. 
In particular, we have characterized the dynamics of a self-diffusiophoretic colloid adsorbed at a fluid-fluid interface for the case in which the colloid is let free to move as well as when the colloid center of mass is kept at a constant position by an external force.
We have found that a rotation of the axis of symmetry of the colloid arises even for fluids with equal viscosity. 
In particular, we have found that the steady state orientation depends on both the opening angle $\theta_\mathrm{o}$ of the catalytic cap and the contact angle $\theta_\mathrm{c}$. 

In order to understand the origin of such a reorientation we have calculated the local mobility by applying locally external, constant forces on a passive colloid. We have extracted the local value of the mobility by taking the ratio between the applied force and the steady-state angular velocity. This local mobility has been used to determine the effective torque acting on the self-diffusiophoretic colloid. 
Interestingly, we have found that, for $\theta_\mathrm{c}=\pi/2$, the mobility matrix predicts quite well the effective torque on a  free colloid as well as on a colloid the center of mass of which is kept spatially  fixed. If  $\theta_\mathrm{c}\neq\pi/2$ we have found good agreement for the case of the fixed colloid, whereas an additional torque arises in the case of a mobile colloid.
Finally, we have discussed the case in which only one fluid contains the reactant. In this case we have  found that the stable steady-state orientation is the one with the cap immersed in the fluid without reactant so that the colloid becomes inactive. 

\appendix
\section{Validation of the numerical scheme\label{sec:homogeneousFluid}}

In order to validate our numerical scheme we characterize the dynamics of a self-diffusiophoretic colloid in a homogeneous bulk fluid. 
In this case, after an initial transient, the colloid attains a steady state in which it moves with a constant velocity $\bm{v} = \bm{e}_{\mathrm{z}} v$. We compare the results of simulations performed for various values of the opening angle $\theta_{\mathrm{o}}$ with analytic results~\cite{Popescu2010}.
\begin{figure}[t!]
    \includegraphics[scale=1]{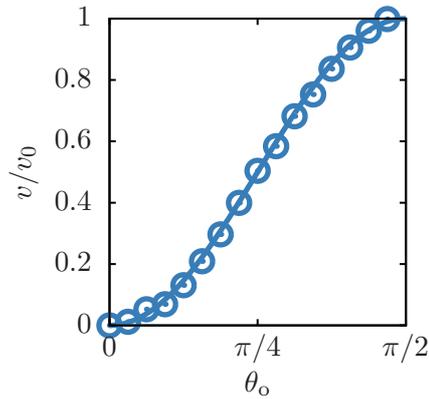}
\caption{\label{fig:velocityCoveringAngle} 
    The velocity $v$ of a self-diffusiophoretic colloid as a function of the opening angle $\theta_{\mathrm{o}}$, normalized by $v_0=\,v(\theta_{\rm{o}}=\pi/2)$ (see Fig.~\ref{fig:velsteadyStates}).
    The symbols are simulation data whereas the solid line provides the analytical  prediction~\cite{Popescu2010}.
}
\end{figure}
Figure~\ref{fig:velocityCoveringAngle} shows the reduced steady state velocity of the colloid as a function of the opening angle. As shown in the figure, the agreement with the analytic results is very good.

\begin{widetext}
\section{Additional results\label{sec:additional}}
In Figs.~\ref{fig:pi4torque},~\ref{fig:pi2torque},~\ref{fig:3pi8torque}, and \ref{fig:pi8torque} the $z$-component of the torques on self-diffusiophoretic colloids with opening angles $\theta_{\mathrm{o}} = \pi / 4,\,\pi/2,\,3\pi/8$, and $\pi/8$,  respectively, are shown as function of the colloid rotation for three contact angles $\theta_{\mathrm{c}}$.

In Figs.~\ref{fig:pi2torque}-\ref{fig:pi8torque} there is still qualitative agreement between the measured torque on the fixed colloid and the torque calculated via Eq.~\eqref{eq:deltaForceIntegral2D}, although the quantitative agreement is weaker.
As before, the torque on the free colloid and the torque on the fixed colloid agree very well for all opening angles if the contact angle is $\theta_{\mathrm{c}} = \pi / 2$, because in this case the lateral motion does not induce a torque on the colloid.

\begin{figure*}[h]
    \includegraphics[scale=0.89]{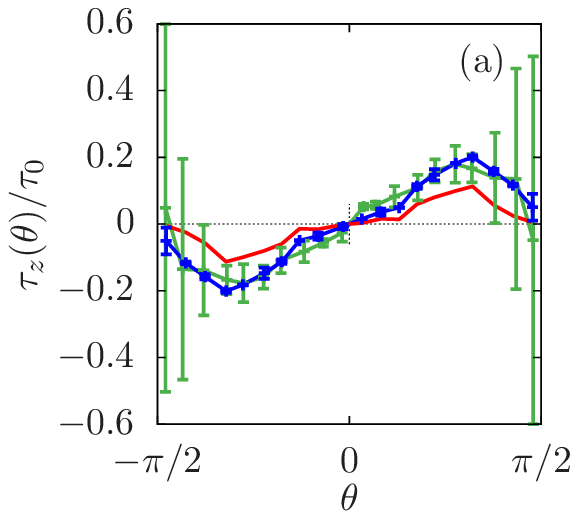}
    \includegraphics[scale=0.89]{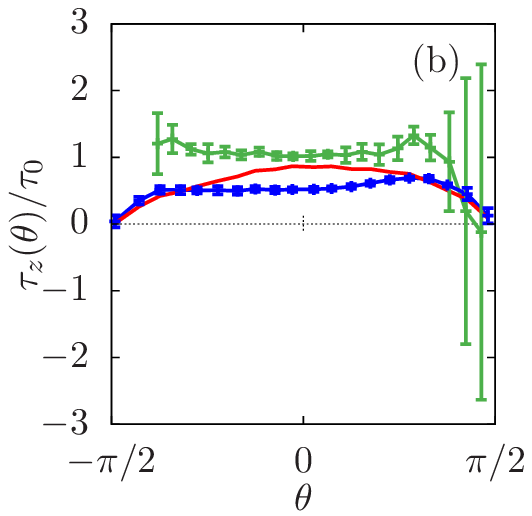}
    \includegraphics[scale=0.89]{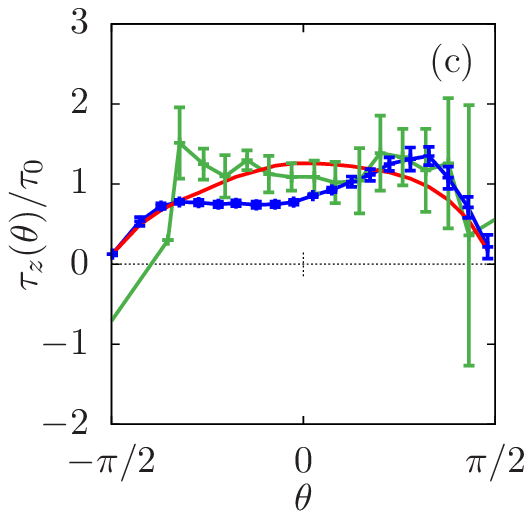}
\caption{
    \label{fig:pi2torque}
    Same as in Fig.~\ref{fig:pi4torque} for $\theta_\mathrm{o}=\pi/2$.
    }
\end{figure*}

\begin{figure*}[htbp!]
    \includegraphics[scale=0.9]{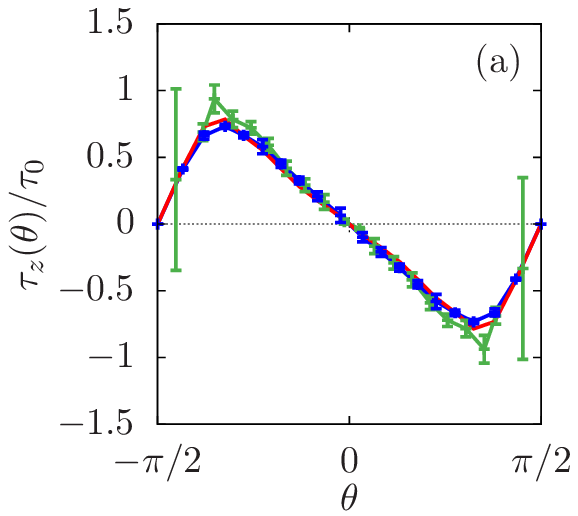}
    \includegraphics[scale=0.9]{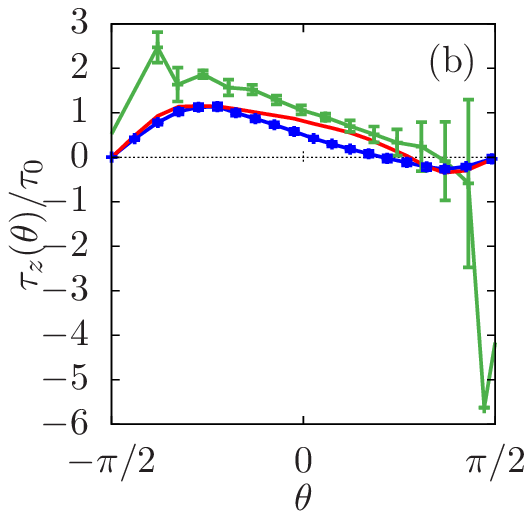}
    \includegraphics[scale=0.9]{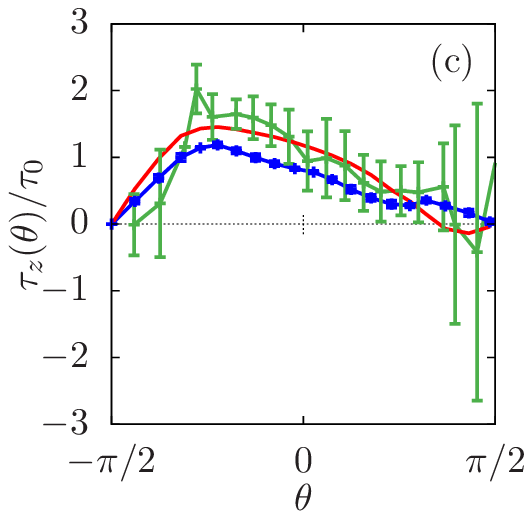}
\caption{
    \label{fig:3pi8torque}
    Same as in Fig.~\ref{fig:pi4torque} for $\theta_\mathrm{o}=3\pi/8$.
    }
\end{figure*}

\begin{figure*}[htbp!]
    \includegraphics[scale=0.9]{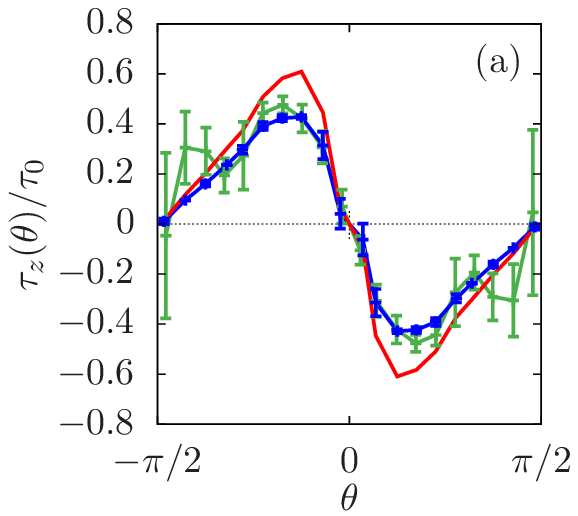}
    \includegraphics[scale=0.9]{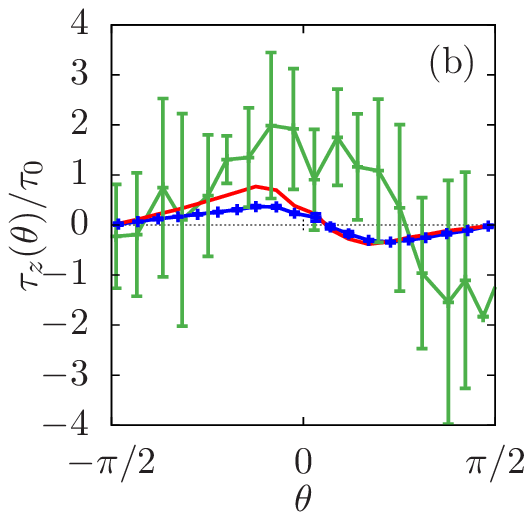}
    \includegraphics[scale=0.9]{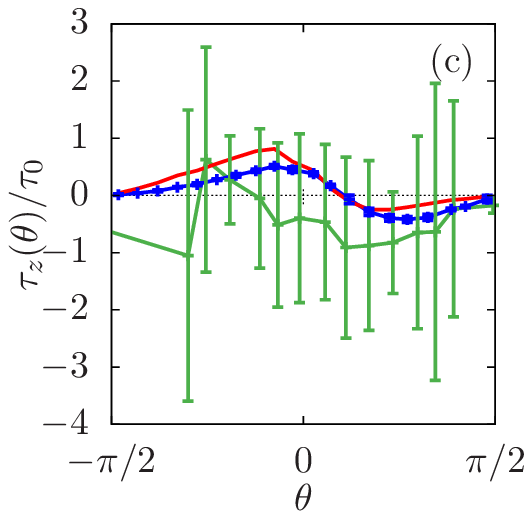}
\caption{
    \label{fig:pi8torque}
    Same as in Fig.~\ref{fig:pi4torque} for $\theta_\mathrm{o}=\pi/8$.
    }
\end{figure*}
\end{widetext}

\bibliography{library}

\end{document}